\documentclass[twocolumn,english,aps,prd,reprint,floatfix,notitlepage,footinbib,preprintnumbers,superscriptaddress,altaffilletter]{revtex4-2}

\usepackage{amsmath,amssymb,amsfonts}
\usepackage{hyperref,breakurl,cleveref,url}
\usepackage{color}

\usepackage{graphicx}
\usepackage[export]{adjustbox}

\usepackage{accents,mathrsfs,mathtools,dsfont}

\renewcommand{\paragraph}[1]{%
    \textit{#1}.---%
}

\def\skip{\vskip1.5pt}
\newcommand\trick[1]{}

\usepackage{enumitem}
\setlist[enumerate]{
    label={},
    leftmargin=2em,
    itemsep=2pt,
    topsep= 2pt,
    partopsep=0pt,
    parsep=0pt,
}

\let\oldeqref\eqref
\renewcommand{\eqref}[1]{Eq.\,\smash{\oldeqref{#1}}}
\newcommand{\eqrefs}[2]{Eqs.\,\smash{\oldeqref{#1}} and \smash{\oldeqref{#2}}}
\newcommand{\eqrefss}[3]{Eqs.\,\smash{\oldeqref{#1}}, \smash{\oldeqref{#2}}, and \smash{\oldeqref{#3}}}

\newcommand{\rcite}[1]{Ref.\,\smash{\cite{#1}}}
\newcommand{\rrcite}[1]{Refs.\,\smash{\cite{#1}}}

\def\mem{\hspace{0.1em}}
\def\hem{\hspace{0.05em}}
\def\nem{\hspace{-0.1em}}
\def\hnem{\hspace{-0.05em}}
\def\hhem{\hspace{0.025em}}
\def\hhnem{\hspace{-0.025em}}
\def\hhhem{\hspace{0.0125em}}
\def\hhhnem{\hspace{-0.0125em}}

\def\blank{{\,\,\,\,\,}}

\def\qiq{{\quad\implies\quad}}

\def\a{\alpha}
\def\b{\beta}
\def\c{{\gamma}}

\def\e{\epsilon}
\def\ve{\varepsilon}
\def\m{\mu}
\def\n{\nu}
\def\r{\rho}
\def\s{\sigma}
\def\k{\kappa}
\def\l{\lambda}
\def\t{\tau}

\def\mtimes{{\mem\times\mem}}
\def\mdot{{\mem\cdot\mem}}

\def\da{{\dot{\a}}}

\newcommand{\wrap}[1]{{\smash{#1}\vphantom{\beta}}}

\def\lsq{{
    \kern-0.037em
    \adjustbox{scale=0.919,valign=c}{$
        {
            \adjustbox{raise=-0.0855em}{$\lfloor$}
            \llap{\reflectbox{\rotatebox[origin=c]{180}{$\lfloor$}}}
        }
    $}
    \kern-0.04em
}}
\def\rsq{{
    \kern-0.04em
    \adjustbox{scale=0.919,valign=c}{$
        {
            \rlap{\reflectbox{\rotatebox[origin=c]{180}{$\rfloor$}}} 
            \adjustbox{raise=-0.0855em}{$\rfloor$}
        }
    $}
    \kern-0.037em
}}

\def\M{{\mathcal{M}}}

\def\mathe{{\scalebox{1.05}[1.01]{$\mathrm{e}$}}}
\def\sprime{{\mathrlap{\smash{{}^\prime}}{\hspace{0.05em}}}}

\def\i{{\iota}}

\def\vex{\vec{x}}
\def\vea{\vec{a}}
\def\vep{\vec{p}}

\newcommand{\cop}[1]{ \mathrlap{c'}\phantom{c}^{\kern0.24em#1} }

\def\S{\Sigma}

\usepackage{tikz}
\usetikzlibrary{calc} 
\usetikzlibrary{shapes.geometric} 
\usetikzlibrary{positioning} 
\usetikzlibrary{fit} 
\usepackage[a]{esvect} 
\tikzset{empty/.style = {inner sep = 0pt, outer sep = 0, minimum size = 0}}
\tikzset{b/.style = {inner sep = 2pt, outer sep = 4pt, minimum size = 12pt}}
\tikzset{w/.style = {inner sep = 1pt, outer sep = 2pt, minimum size = 12pt, anchor = west}}
\tikzset{s/.style = {inner sep = 2.5pt, outer sep =2.5pt, minimum size = 1pt, font = \small}}

\definecolor{sky}{RGB}{144,187,231}
\definecolor{OxyRed}{RGB}{190,70,62}
\definecolor{NitroBlue}{RGB}{91,122,239}
\definecolor{HydrogenLight}{RGB}{245,250,252}

\tikzset{
	line/.style = {draw, line width = 1.1pt, line cap = round, rounded corners = 0.2pt},
	bine/.style = {draw, line width = 1.1pt, line cap = round, rounded corners = 0.0pt, dotted, color=OxyRed},
	dine/.style = {draw, line width = 1.4pt, line cap = round, rounded corners = 0.0pt, double},
	D/.style = {below = -1.1pt, font = \footnotesize\bfseries\sffamily},
	U/.style = {above = -1.2pt, font = \footnotesize\bfseries\sffamily},
	L/.style = {left,  font=\footnotesize},
	R/.style = {right, font=\footnotesize},
	X/.style = {circle, draw=black, fill=HydrogenLight, inner sep=0pt, outer sep=0pt, minimum size=4.5pt, line width=1.1pt},
	Y/.style = {circle, draw=black, fill=NitroBlue, inner sep=0pt, outer sep=0pt, minimum size=4.5pt, line width=1.1pt}
}

\usepackage[export]{adjustbox}

\newcommand{\bb}[1]{\bigg(\,{#1}\,\bigg)}
\newcommand{\BB}[1]{\Big(\,{#1}\,\Big)}

\def\mtimes{{\mem\times\mem}}
\def\mdot{{\mem\cdot\mem}}
\def\mwedge{{\mem\wedge\mem\hhem}}
\def\swedge{{\mem{\wedge}\,}}

\def\mplus{{\mem+\mem}}

\def\mtimes{{\mem\times\hem}}
\def\mdot{{\mem\cdot\mem}}
\def\mt{{\mem\times}}

\def\O{\mathcal{O}}

\def\rambda{\bar{\lambda}}

\begin{document}

\title{
	Note on the Kerr Spinning-Particle Equations of Motion
}

\author{Joon-Hwi Kim}
\affiliation{Walter Burke Institute for Theoretical Physics, California Institute of Technology, Pasadena, CA 91125}

\begin{abstract}
	We implement a probe counterpart of Newman-Janis algorithm,
	which Wick rotates the all-orders geodesic deviation equation 
	into a part of exact spinning-particle equations of motion.
	Consequently,
	the gravitational dynamics of the Kerr black hole in its point-particle effective theory
	is completely constrained in the self-dual sector
	for a hidden symmetry,
	implying
	the spin exponentiation of same-helicity gravitational Compton amplitudes to all multiplicities.
\end{abstract}

\preprint{CALT-TH 2025-006}


\bibliographystyle{utphys-modified}

\renewcommand*{\bibfont}{\fontsize{8}{8.5}\selectfont}
\setlength{\bibsep}{1pt}

\maketitle

\paragraph{Introduction}%
Suppose a free-falling probe particle in space,
followed by an observer.
Their separation
is famously described by 
the geodesic deviation equation.
Yet, typical textbooks
describe this equation only in
the limit of infinitesimal separation.
The exact equation
valid to all orders in \textit{separation} and \textit{curvature}
was investigated by Vines \cite{Vines:2014oba}.
Recently,
we have revisited this topic
from an alternative formalism \cite{gde}.

It seems that
this foundational subject of geodesic deviation
is intimately tied to
a pressing problem
in contemporary gravitational physics:
the exact dynamics of the Kerr black hole as a probe object.

Suppose a Kerr black hole
in the presence of mild gravitational perturbations;
it may travel through macroscopic gravitational waves
or constitute an inspiral-phase binary system.
It is well-established
\cite{Goldberger:2004jt,Porto:2005ac,Levi:2015msa,Porto:2016pyg,Levi:2018nxp,Kalin:2020mvi}
that
the black hole
can be treated as a relativistic spinning point particle
in the \textit{effective theory} sense,
provided separation of scales.
In this 
effective description,
the persistent problem
has been
determining
the exact equations of motion (EoM) for the Kerr black hole
in the probe limit,
to all orders in the so-called \textit{spin length}
(spin per mass, encoding the ring radius)
and \textit{curvature}.
%
%
%
%
%
At the linear order in curvature,
the EoM
have been completely fixed
to all orders in spin length
\cite{ambikerr1,gmoov}.
Frustratingly,
ambiguities arise
from the very quadratic order in curvature:
in layman's terms,
physicists in this 21\textsuperscript{st} century
do not know the \smash{``$\vec{F} = m\vec{a}$\mem''} of
spinning black holes.
This agony has gained a considerable attention
lately
and has been scrutinized
in terms of scattering amplitudes
\cite{ahh2017,%
Guevara:2018wpp,Guevara:2019fsj,chkl2019,aho2020,%
Johansson:2019dnu,Aoude:2020onz,Lazopoulos:2021mna,
Bern:2022kto,Aoude:2022trd,
Falkowski:2020aso,Chiodaroli:2021eug,
Ochirov:2022nqz,Cangemi:2022bew,
fabian1,fabian2,zihan23,Saketh:2023bul,
Cangemi:2022abk,Alessio:2025nzd
}.

The widespread expectation is that 
the Kerr black hole will exhibit the \textit{simplest} EoM
among all massive spinning objects
in four dimensions,
in light of its simplicity
uncovered in a modern analysis \cite{ahh2017,Guevara:2018wpp,Guevara:2019fsj,chkl2019,aho2020%
}
that traces back to the Newman-Janis algorithm (NJA) \cite{Newman:1965tw-janis}.
The NJA,
introduced by the very article
``Note on the Kerr Spinning-Particle Metric'' in 1965 \cite{Newman:1965tw-janis},
is a method that
derives the Kerr solution
by displacing the Schwarzschild solution
along a complexified direction
in a sense.
Crucially, this
reinterprets spin as an imaginary deviation \cite{newman1974curiosity,newman1974collection}.
Due to \rcite{nja},
the NJA now holds the status of 
a rigorous derivation of the Kerr solution
with a definite origin.

This article shows that
a unique answer exists for the simplest spinning-particle EoM
to all orders in spin and curvature,
in self-dual spacetimes.
This Wick rotates the \textit{deviation vector} in the exact geodesic deviation equation
to the \textit{spin length pseudovector} in exact spinning-particle EoM,
as an incarnation of NJA
at the probe level.

Our proposal is distinguished in terms of a hidden symmetry
in the presence of Killing-Yano tensors.
In perturbation theory,
it implies
the property known as
spin exponentiation \cite{ahh2017,Guevara:2018wpp,Guevara:2019fsj,chkl2019,aho2020,Johansson:2019dnu,Aoude:2020onz,Lazopoulos:2021mna}
at all graviton multiplicities.

\skip
\paragraph{Schwarzschild EoM}%
The free-fall EoM of a scalar probe
is given in the first-order formulation as
\begin{align}
	\label{GDE.z}
	p^{\m\sprime}
	\mem=\mem
		\frac{dz^{\m\sprime}}{d\t}
	\,,\quad
	\frac{Dp^{\m\sprime}}{d\t}
	=\,
		0
	\,,
\end{align}
where $D$ denotes the covariant differential.
$z^{\m\sprime}$ describes the position,
while $p^{\m\sprime}$ describes the momentum.
We have put primes on the indices
with hindsight.

Now introduce an arbitrary observer at position $x^\m$.
\rcite{gde} shows that
\eqref{GDE.z} is equivalent to
\begin{align}
	\label{GDE}
	p^\m
	=\,
		\a^\m{}_\n\mem \frac{dx^\n}{d\t}
		+ \b^\m{}_\n\mem \frac{Dy^\n}{d\t}
	\,,\quad
	\frac{Dp^\m}{d\t}
	=\,
		\Omega^\m{}_\n\mem p^\n
	\,,
\end{align}
where
$\a^\m{}_\n$,
$\b^\m{}_\n$,
and
$\Omega^\m{}_\n$
are tensors
composed of
the Riemann tensor $R^\m{}_{\n\r\s}(x)$, its covariant derivatives,
and the vector $y^\m$,
etc.,
all transforming tensorially at $x$.
Also, $\Omega_{\m\n} = -\Omega_{\n\m}$
so that $p^2$ is conserved.

The idea is to construct a geodesic joining between $x$ and $z$,
whose tangent at $x$ is $y^\m$:
\begin{subequations}
\label{gds}
\begin{align}
	\label{z}
	z^{\m\sprime}
	\mem=\mem
		\delta^{\m\sprime}{}_\m\mem
		\BB{
			x^\m + y^\m - \tfrac{1}{2}\mem \Gamma^\m{}_{\r\s}(x)\mem y^\r y^\s 
			+ \O(y^3)
		}
	\,.
\end{align}
In the well-known Synge formalism
\cite{Ruse:1931ht,Synge:1931zz,Synge:1960ueh,Poisson:2011nh},
the parallel propagator
about the geodesic path
is given by
\begin{align}
	\label{W}
	W^{\m\sprime}{}_\n
	\mem=\mem
		\delta^{\m\sprime}{}_\n
		-
			\delta^{\m\sprime}{}_\m\mem
			\Gamma^\m{}_{\n\r}(x)\mem y^\r
		+ \O(y^2)
	\,,
\end{align}
\end{subequations}
so a vector $p^\m$ based at $x$ is parallel-transported to $z$ as
$p^{\m\sprime} =$ $ W^{\m\sprime}{}_\n\mem p^\n$.
Especially,
the geodesic's tangent vector at $z$ is
$y^{\m\sprime} = W^{\m\sprime}{}_\n\mem y^\n$.
Multiplying 
\eqref{GDE.z}
by
the inverse
$W^\m{}_{\m'}$ of $W^{\m\sprime}{}_\m$
re-covariantizes tensors at $z$
with respect to the point $x$,
hence deriving \eqref{GDE}.

\rcite{gde} provides all-orders formulae for
$\a^\m{}_\n$, $\b^\m{}_\n$, and $\Omega^\m{}_\n$
explicitly as power series in $y$:
\begin{align}
\begin{split}
	\label{alpha}
	\kern-0.1em
	\a^\m{}_\n ={}
	&
		\delta^\m{}_\n 
		+ \tfrac{1}{2}\mem y^\r y^\s R^\m{}_{\r\s\n}(x)
		+ \tfrac{1}{6}\mem y^\r 
		y^\s y^\k R^\m{}_{\r\s\n;\k}(x)
	\\
	&
		+ \tfrac{1}{24}\mem y^\r y^\s y^\k y^\l R^\m{}_{\r\s\n;\k\l}(x)
	\\
	&
		+ \tfrac{1}{24}\mem y^\r y^\s y^\k y^\l R^\m{}_{\r\s\xi}(x)\hem R^\xi{}_{\k\l\n}(x)
		+ \O(y^5) 
	\,,
	\kern-0.1em
\end{split}
\end{align}
for instance.
Note the nonlinearity in curvature.

\eqref{GDE}
is the first-order EoM of
a free-falling probe
as seen by an arbitrary observer.
The dynamical variables are
$y^\m$ and $p^\m$,
while $x^\m$ is a nondynamical reference.
The second-order version is worked out in \rcite{gde}.

\newpage

\eqref{GDE}
characterizes the exact dynamics of a minimally coupled scalar probe,
which can model a Schwarzschild black hole in its point-particle effective theory.

\skip
\paragraph{``Derivation'' of Kerr EoM}%
Now we implement a three-step procedure that derives
a distinct class of exact spinning-particle EoM
in four dimensions.

Firstly,
take the Schwarzschild EoM in \eqref{GDE}
for an imaginary (Wick-rotated) deviation:
\begin{align}
	\label{wick}
	y^\m 
		\,\,\mapsto\,\,
	iy^\m
	\,.
\end{align}
The result reads
\begin{subequations}
\label{heaven}
\begin{align}
	\label{heaven.x}
	p^\m
	&=\,
		\a^\m{}_\n\mem \frac{dx^\n}{d\t}
		+ i \b^\m{}_\n\mem \frac{Dy^\n}{d\t}
	\,,\\
	\label{heaven.p}
	\frac{Dp^\m}{d\t}
	&=\,
		\Omega^\m{}_\n\mem p^\n
	\,,
\end{align}
with the understanding that
the expressions for
$\a^\m{}_\n$,
$\b^\m{}_\n$,
and $\Omega^\m{}_\n$
should change accordingly
by analytic continuation of power series:
for instance,
$\a^\m{}_\n$ in \eqref{heaven.x}
equals
$
	\delta^\m{}_\n 
	- \frac{1}{2}\mem y^\r y^\s R^\m{}_{\r\s\n}(x) 
	- \frac{i}{6}\mem y^\r y^\s y^\k R^\m{}_{\r\s\n;\k}(x) 
	+ \O(y^4)
$.

Secondly,
stipulate an additional pair of equations,
\begin{align}
	\label{heaven.y}
	\frac{Dy^\m}{d\t}
	\,&=\,
		\Omega^\m{}_\n\mem y^\n
	\,,
\end{align}
\end{subequations}
which demands
that the precession of $y^\m$
is synchronized with that of
the momentum
$p^\m$ in \eqref{heaven.p}.

Let us pause for a second to clarify the interpretations.
The Wick rotation in \eqref{wick}
implements the idea of NJA
at the level of probe EoM,
so $y^\m$ is now interpreted as the spin length pseudovector.
Its definition reads
\begin{align}
	\label{ydef}
	y^\m 
	\,=\,
		-{*}S^{\m\n} p_\n / p^2
	\,,
\end{align}
where $S^{\m\n}$ is the spin angular momentum tensor.
\eqref{ydef} is equivalent to imposing
$S^{\m\n} = \varepsilon^{\m\n\r\s} y_\r p_\s$
and $p \mdot\hnem y = 0$.

Consequently,
all of $x^\m$, $p^\m$, and $y^\m$
had to be interpreted as dynamical variables
of a spinning particle,
requiring a total of $12$ equations.
The additional stipulation in \eqref{heaven.y}
was exactly for this purpose.
Physically, such a synchronization of orbital ($p^\m$)
and spin ($y^\m$) precessions
will encode
a spinning/complexified analog of equivalence principle 
proposed in \rcite{sst-asym}.

Finally,
we enter the last step of our procedure.
The $12$ equations in \eqrefss{heaven.x}{heaven.p}{heaven.y}
together defines well-posed spinning-particle EoM,
iff the variables $x^\m, p^\m, y^\m$ are complexified.
This is because the Wick rotation in \eqref{wick}
induces imaginary units inside the tensors
$\a^\m{}_\n$, $\b^\m{}_\n$, $\Omega^\m{}_\n$.
Therefore,
\eqref{heaven} describes
the EoM of a spinning particle in \textit{complexified spacetime}.

Thus, we propose that
each imaginary unit in \eqref{heaven}
is eliminated by trading off with
a Hodge star $*$ on a Riemann tensor,
which is justified if the complexified spacetime is taken as \textit{self-dual}:
$i \mapsto *$ via
$i\hem R_{\m\n\r\s} = {*}R_{\m\n\r\s}$.
In this manner, \eqref{heaven}
is uplifted to a set of completely real equations
that dictates the time evolution of $x^\m,p^\m,y^\m$ as real variables.
Details will be elaborated later.

Parity property of $y^\m$ implies that
the linear-in-curvature part of \eqref{heaven}
reproduces the unity 
\cite{Newman:1965tw-janis,Newman:1973yu,janis1965structure}
of spin-induced multipole moments
of the Kerr black hole,
if uplifted in this precise fashion.
This provides a justification.
Intuitively, we have turned the localized orbital momentum
in geodesic deviation
(yielding unity of mass multipoles)
into a spin angular momentum.

In sum, we have proposed a probe counterpart of NJA
that uniquely derives
a well-posed instance of spinning-particle EoM
in \textit{heaven},
which will then be readily uplifted to \textit{earth}.
Here, 
we have used technical terminologies coined by 
Newman \cite{shaviv1975general,Newman:1976gc} and Pleba\'nski \cite{plebanski1975some,Plebanski:1977zz}:
\begin{align}
\begin{split}
	\label{newman-plebanski}
	\textsc{Heaven}
	\,&=\,
	\textsc{Self-Dual Spacetime}
	\,,\\
	\textsc{Earth}
	\,&=\,
	\textsc{Real Spacetime}
	\,.
\end{split}
\end{align}
To clarify,
heaven is a complex-analytic four-manifold with a holomorphic metric,
while
earth is a real-analytic four-manifold with a pseudo-Riemannian metric.

Remarkably,
the following hold for \eqref{heaven} in heaven:
\begin{enumerate}
\item 
	(\textsc{a})
		The linear-in-curvature part of
		\eqref{heaven} describes the established linear coupling
		for Kerr.
\item 
	(\textsc{b})
		The dynamics due to \eqref{heaven}
		features exact (all-orders) conserved quantities
		in the presence of Killing vectors and Killing-Yano tensors.
\item 
	(\textsc{c})
		\eqref{heaven} admits a Lagrangian formulation.
\end{enumerate}
In these senses,
\eqref{heaven}
is an exact and consistent nonlinear completion
of the linearized Kerr spinning-particle EoM,
characterized by hidden symmetry.
Hence we propose that
\eqref{heaven}
is \textit{the} exact Kerr spinning-particle EoM
in self-dual spacetimes.
Below, we refer to \eqref{heaven}
as the \textit{heavenly Kerr spinning-particle EoM}.

\skip
\paragraph{Hidden Symmetry}%
Let us begin with the proof of
proposition (\textsc{b}).
First of all,
recall the equivalence between \eqrefs{GDE.z}{GDE}
via \eqref{gds}.
Due to the Wick rotation in \eqref{wick},
\eqref{gds} is now redefined as
\begin{subequations}
\label{igds}
\begin{align}
	\label{z.wick}
	z^{\m\sprime}
	&=\mem
		\delta^{\m\sprime}{}_\m\mem
		\BB{
			x^\m + iy^\m + \tfrac{1}{2}\mem \Gamma^\m{}_{\r\s}(x)\mem y^\r y^\s 
			+ \O(y^3)
			\nem
		}
	\,,
	\kern-0.3em\\
	\label{W.wick}
	\kern-0.2em
	W^{\m\sprime}{}_\n
	&=\mem
		\delta^{\m\sprime}{}_\n
		- i\mem
			\delta^{\m\sprime}{}_\m\mem
			\Gamma^\m{}_{\n\r}(x)\mem y^\r
		+ \O(y^2)
	\,.
\end{align}
\end{subequations}
\eqref{z.wick}
describes a complexified geodesic
for which the complexified parallel propagator is \eqref{W.wick}.
These are well-established ideas in differential geometry:
adapted complex structure
\cite{guillemin1991grauert,guillemin1992grauert,lempert1991global,szHoke1991complex,halverscheid2002complexifications,aguilar2001symplectic,burns2000symplectic,hall2011adapted}.
See \rcite{gmoov} for a reinvention.

Via \eqref{igds},
\eqref{heaven} is equivalent to
\begin{align}
	\label{heaven.z}
	p^{\m\sprime}
	\mem=\mem
		\frac{dz^{\m\sprime}}{d\t}
	\,,\quad
	\frac{Dp^{\m\sprime}}{d\t}
	\mem=\mem
		0
	\,,\quad
	\frac{Dy^{\m\sprime}}{d\t}
	\mem=\mem
		0
	\,,
\end{align}
where $p^{\m\sprime} = W^{\m\sprime}{}_\m\mem p^\m$
and
$y^{\m\sprime} = W^{\m\sprime}{}_\m\mem y^\m$
via \eqref{W.wick}.
Again, multiplying by $W^{\m\sprime}{}_\m$
yields \eqref{heaven.z}
from \eqref{heaven}.

By the very construction of the probe NJA,
the first two equations in
\eqref{heaven.z}
are simply 
the free-fall equations in \eqref{GDE.z}
with the revised definitions 
of $z^{\m\sprime}$ and $p^{\m\sprime}$
due to \eqref{igds}.
Consequently,
a part of 
the heavenly Kerr spinning-particle EoM
is literally
the Schwarzschild scalar-particle EoM,
albeit imaginary-shifted.

The last equation in \eqref{heaven.z},
on the other hand,
is a remarkable restatement of the orbit-spin synchronization.
It is amusing to see 
how the black hole's
spin precession parameters
such as the
gravimagnetic ratio $\k {\,=\,} 1$ \cite{Yee:1993ya,khriplovich1989particle,Khriplovich:1997ni}
are precisely reformulated 
as parallel transport (covariant constancy) of $p^{\m\sprime}, y^{\m\sprime}$
along the complex worldline $z^{\m\sprime}$.

Next,
recall the following well-known facts
\cite{Carter:1968ks,yano1952some,Walker:1970un,Hughston:1972qf,hughston1973spacetimes,jeffryes1984space,Jezierski:2005cg,Nozawa:2015qea,hansen2014killing,Penrose:1973um,floyd1973dynamics,marck1983solution}
about free-fall motion
with position $x^\m$ and momentum $p^\m$.
First, a vector field $X^\m(x)$
is Killing if $X^{\m;\r}(x) = X^{[\m;\r]}(x)$,
which implies constancy of
$p_\m\hem X^\m(x)$.
Second, a two-form $Y_{\m\n}(x)$ 
is Killing-Yano 
if $Y_{\m\n;\r}(x) = Y_\wrap{[\m\n;\r]}(x)$,
which implies covariant constancy of
$Y^\m{}_\n(x)\mem p^\n$.
In turn, the scalar combination
$(\hhem Y(x)\hem p)^2 = -p_\m Y^\m{}_\r(x)\mem Y^\r{}_\n(x)\mem p^\n$
is strictly constant
along the worldline.

Via complexification, these facts are immediately 
inherited to the Kerr spinning particle:
\begin{align}
	\label{pX}
	\frac{d}{d\t}\mem
	\BB{
		p_{\m'} X^{\m\sprime}(z)
	}
	\mem&=\,
		p_{\m'} X^{\m\sprime}{}_{;\r'}(z)\, p^{\r\sprime}
	\,=\,
		0
	\,,\\
	\label{Yp}
	\frac{D}{d\t}\mem
	\BB{
		Y^{\m\sprime}{}_{\n'}\hnem(z)\mem p^{\n\sprime}
	\mem}
	\mem&=\,
		Y^{\m\sprime}{}_{\n';\r'}\hnem(z)\mem p^{\n\sprime} p^{\r\sprime}
	\,=\,
		0
	\,.
\end{align}
Meanwhile, the spinning particle features
a yet another variable
that is covariantly constant:
$y^{\m\sprime}$ in \eqref{heaven.z}.
Consequently,
$p^{\m\sprime}$,
$Y^{\m\sprime}{}_{\n'}\hnem(z)\mem p^{\n\sprime}$,
and
$y^{\m\sprime}$
are
parallel-transported altogether along the complex worldline $z^{\m\sprime}$,
so
any scalar product between them
is strictly constant.

In this way, it holds true that
the following
are exact, meaning all-orders in spin and curvature,
conserved quantities
of the Kerr spinning particle:
\begin{align}
	\label{Q}
	\mathds{Q}
	\,&=\,
		p_{\m'} X^{\m\sprime}(z)
	\,,\\
	\label{R}
	\mathds{R}
	\,&=\,
		p_{\m'} Y^{\m\sprime}{}_{\n'}\hnem(z)\mem y^{\n\sprime}
	\,,\\
	\label{C}
	\mathds{C}
	\,&=\,
		- p_{\m'} Y^{\m\sprime}{}_{\r'}\hnem(z)\mem Y^{\r\sprime}{}_{\n'}\hnem(z)\mem p^{\n\sprime}
	\,.
\end{align}
\eqrefs{R}{C} are
all-orders completions of the R\"udiger \cite{rudiger1981conserved,rudiger1983conserved} and Carter \cite{Carter:1968ks} constants,
respectively.

Explicitly,
\eqref{Q} can be re-covariantized 
at $x$ as
\begin{align}
\begin{split}
	\label{pX.split}
	p_\m\hem W^\m{}_{\m\sprime}\, X^{\m\sprime}(z)
	\mem=\,
		\sum_{\ell=0}^\infty\,
			\frac{i^\ell}{\ell!}\,
				p_\m\mem 
				X^\m{}_{;\r_1\cdots\r_\ell}(x)\mem
				y^{\r_1} {\cdots\hem} y^{\r_\ell}
	\,,
	\nem
\end{split}
\end{align}
yielding an infinite expansion in the spin length.
When written in this form,
its conservation
becomes quite nontrivial to check:
the use of identities regarding $\a^\m{}_\n$, $\b^\m{}_\n$, $\Omega^\m{}_\n$
is mandated
at each order.
This highlights the power of the spin-resummed formulation
facilitated by \eqref{heaven.z}.
The same analysis applies to \eqrefs{R}{C} as well.

Furthermore,
heaven is stipulated to be self-dual,
so its anti-self-dual spinor bundle is flat.
As a result, the anti-self-dual Pleba\'nski two-forms
\cite{Plebanski:1977zz,capovilla1991self}
are covariantly constant:
$(\S^a)_{\m\n}$ for $a {\,=\,} 1,2,3$.
These serve as Killing-Yano tensors
\cite{Gibbons:1987sp,Nozawa:2015qea}
and define a quaternionic structure
as
$
    (\S_a)^\m{}_\r\hem (\S_b)^\r{}_\n
    = -\delta_{ab}\mem \delta^\m{}_\n
    + \ve^c{}_{ab}\mem (\S_c)^\m{}_\n
$
\cite{Gibbons:1987sp,Atiyah:1978wi,Nozawa:2015qea}.
Consequently,
we now identify
$p^{\m\sprime}$,
$Y^{\m\sprime}{}_{\n'}\hnem(z)\mem p^{\n\sprime}$,
$(\Sigma_a)^{\m\sprime}{}_{\n'}\hnem(z)\mem p^{\n\sprime}$,
$y^{\m\sprime}$
being covariantly constant altogether,
so any scalar product between them is a constant of motion,
such as
\begin{align}
	\label{K}
	\mathds{K}_a
	\,=\,
		p_{\m'} (\Sigma_a)^{\m\sprime}{}_{\r'}\hnem(z)\mem Y^{\r\sprime}{}_{\n'}\hnem(z)\mem p^{\n\sprime}
	\,.
\end{align}

A physically relevant example
is the self-dual Taub-Newman-Unti-Tamburino (NUT) background,
which describes a gravitational instanton
\cite{hawking1977gravitational,Gibbons:1978tef}.
It is now well-understood 
\cite{nja,note-sdtn,Newman:1973yu,Newman:2002mk,Gross:1983hb,Ghezelbash:2007kw,Crawley:2021auj,gabriel1,gabriel2,Adamo:2023fbj}
to an explicit extent \cite{nja,note-sdtn}
that this background
is the self-dual sector of the Kerr background,
such that the same Killing-Yano tensor is shared \cite{note-sdtn,nja}.
By using the Kerr-Schild metric provided by \rcite{note-sdtn},
\eqref{K} evaluates to
$\vec{\mathds{K}} = \vep \mt (\vex\mtimes\vep) - ip_0\mem (\vex\mt\vep)$
\footnote{
	Also, the three-vector index of $\mathds{K}_a$ is global
	since we gauge-fixed the anti-self-dual spinor bundle
	by zeroing the spin connection coefficients.
},
which
amusingly
reincarnates
the Laplace-Runge-Lenz vector
(cf.\:\rrcite{Gibbons:1986df,Gibbons:1986hz,Gibbons:1987sp,Feher:1986ib,Feher:1988th,Cordani:1989ip,Iwai:1992zu,Guevara:2023wlr}).

By incorporating time translation and rotation invariances as well,
we establish that
the motion of the Kerr spinning-particle
in the self-dual sector of the Kerr background
is \textit{superintegrable},
with hidden symmetry $\mathrm{so}(4)$
\footnote{
	with suitable complexification
}:
that of
Kepler problem.
This is a new result on the all-orders dynamics of
the Kerr-Kerr system.
Previous works have truncated the probe's spin order \cite{Compere:2023alp,Jakobsen:2021susysky,Gibbons:1993ap,Akpinar:2025tct}
or Newman-Janis shifted only the background \cite{Guevara:2023wlr,Guevara:2024edh}.

To summarize,
we have shown that
exact symmetries arise
for our proposed Kerr spinning particle
by the very structure of the probe NJA.
The conserved quantities of the spinning probe
can arise as direct Newman-Janis shifts of
the conserved quantities of the scalar probe.
For the explicit example of physical relevance,
we have found a sufficient number of functionally independent conserved quantities,
establishing superintegrability.

\skip
\paragraph{Lagrangian Formulation}%
Next, we prove proposition (\textsc{c}).
We find an explicit first-order worldline action,
\begin{align}
	\label{Lz}
	\int\mem
		p_{\m'}\mem \BB{
			dz^{\m\sprime} 
			- i\mem Dy^{\m'}
			+ {*}\Theta^{\m\sprime}{}_{\n'}\mem y^{\n\sprime}
		\,}
		- \mathscr{C}
	\,,
\end{align}
where we have omitted constraint terms ($\mathscr{C}$)
suitably
encoding the conditions such as $p^2 = -m^2$ and $p \mdot\hnem y = 0$.
Computation shows that
the saddle of \eqref{Lz}
is precisely \eqref{heaven.z};
see the appendix for details.

In \eqref{Lz},
$\Theta^{\m\sprime}{}_{\n'}$
describes the ``infinitesimal angle''
as a one-form,
coupled to $S^{\n\sprime}{}_{\m'\nem} = {*}(y \mwedge p)^{\n\sprime}{}_{\m\sprime}$
to realize the kinetic term for spin.
The covariant exterior derivative $D$ acts as
$D\Theta^{\m\sprime}{}_{\n'}
= \frac{1}{2}\mem R^{\m\sprime}{}_{\n'\r'\s'}(z)\mem dz^{\r\sprime} \swedge dz^{\s\sprime}
$,
implying
\begin{align}
	\label{D=0}
	D\mem
	\BB{
		- i\mem Dy^{\m'}
		+ {*}\Theta^{\m\sprime}{}_{\n'}\mem y^{\n\sprime}
	\,}
	\,=\,
		0
\end{align}
for the self-duality.
Via \eqref{D=0},
\eqref{Lz} is equivalent to
\begin{align}
	\label{L}
	\int\mem
		p_\m\mem \bb{\hem
		\begin{aligned}[c]
		&
			\a^\m{}_\n\mem dx^\n
			+ i\mem \b^\m{}_\n\mem Dy^\n
		\\
		&
			- i\mem Dy^\m
			+ {*}\Theta^\m{}_\n\mem y^\m
		\end{aligned}
		}
		- \mathscr{C}
	\,,
\end{align}
which re-covariantizes a one-form
at the point $x$ \cite{gde}.
Both
\eqrefs{Lz}{L}
retrieve the action of a free massive spinning particle
\cite{Hanson:1974qy,ambikerr0}
in the flat limit.

\eqref{Lz} is a complexified action
dictating the evolution of
complexified variables $z^{\m\sprime}$, $p^{\m\sprime}$, $y^{\m\sprime}$.
When converted to \eqref{L},
it can be uplifted to a purely real action
dictating the evolution of real variables
$x^\m$, $p^\m$, $y^\m$.

\paragraph{Linearized Coupling}%
Finally, we prove proposition (\textsc{a})
by showing that
dropping all terms nonlinear in the curvature
in \eqref{L}
reproduces the well-known 
Levi-Steinhoff action \cite{Levi:2015msa}
up to iterating EoM.
The only important term is $p_\m\mem \a^\m{}_\n\mem dx^\n$,
as $\beta^\m{}_\n - \delta^\m{}_\n = \O(R^1y^2)$ \cite{gde} and
$Dy^\m\hnem/d\t = \Omega^\m{}_\n\mem y^\n = \O(R^1y^2)$
on EoM.
As is glimpsed in \eqref{alpha},
the $\O(R^1)$ part of $\a^\m{}_\n$
in heaven
is given by \cite{gde}
\begin{align}
	\label{alphaR1}
	\delta^\m{}_\n
	\,+\,
	\sum_{\ell=2}^\infty\,
		\frac{1}{\ell!}\,
			{*}^\ell R^\m{}_{\k_1\k_2\n;\k_3;\cdots;\k_\ell}(x)\,
			y^{\k_1} \cdots y^{\k_\ell}
	\,,
\end{align}
after the Wick rotation in \eqref{wick}.
Via $dx^\m\hnem/d\t = p^\m + \O(R^1y^2)$
and $S^{\m\n} = \ve^{\m\n\r\s} y_\r p_\s$,
the relevant parts of the action boil down to
$\int p_\m\mem \a^\m{}_\n\mem p^\n - \frac{1}{2}\mem S^{\m\n} {*}\Theta_{\m\n} - \mathscr{C}$,
reproducing the
Levi-Steinhoff action \cite{Levi:2015msa}
for the Kerr multipole coefficients $C_\ell = 1$
\cite{ahh2017,Guevara:2018wpp,Guevara:2019fsj,chkl2019,aho2020}
upon
plugging in \eqref{alphaR1}. 


In fact, the unity $C_\ell {\:=\,} 1$ of multipole coefficients
can be shown directly
from the EoM
by matching  
$\Omega^\m{}_\n$ in \eqref{heaven}
with
the universal template for spin precession EoM \cite{ambikerr1}
extending the Mathisson-Papapetrou-Dixon  \cite{Mathisson:1937zz,Papapetrou:1951pa,Dixon:1970zza}
equations.
Invariantly, it is verified by the spin exponentiation for one graviton
(see \eqref{spin-exp}).

\skip
\paragraph{Spin Exponentiation at All Multiplicities}%
Finally, let us explore all orders in perturbation theory:
\begin{enumerate}
\item 
	(\textsc{d})
		Positive-helicity graviton Compton amplitudes
		due to the action in \eqref{Lz}
		exhibit spin exponentiation at all multiplicities.
\end{enumerate}
No physical explanation has existed to
this property
to our best knowledge,
although its consistency at the amplitudes level has been known \cite{Johansson:2019dnu,Aoude:2020onz,Lazopoulos:2021mna}.
In our case,
the structure of the probe NJA
is highly constrained for the hidden symmetries in the self-dual sector,
whose implication is spin exponentiation at all multiplicities.

To prove proposition (\textsc{d}),
we expand the gravitational fields around Minkowski background,
yet after switching to the tetrad formalism.
In tetrad formalism, the constraints and Hamiltonian (put in $\mathscr{C}$ in \eqref{Lz}) take the same form as in the free theory.
Also, the spin connection describes antisymmetric indices as 
$\gamma^{AB}{}_\r = - \gamma^{BA}{}_\r$,
on which the Hodge star 
acts to yield $+i$ in heaven.

Amusingly,
this shows that the coupling to the spin connection
identically vanish in \eqref{Lz},
consistently with \eqref{D=0}.
Thus the only graviton coupling is via the coframe perturbation,
$h^A{}_\m = e^A{}_\m - \delta^A{}_\m$:
the interaction Lagrangian describes a single term,
$p_{A'}\mem h^{A\sprime}{}_{\m'}(z)\mem \dot{z}^{\m\sprime}$.
This is identical to the interaction Lagrangian 
$p_A\mem h^A{}_\m(x) \dot{x}^\m$
of
the minimally coupled scalar particle
in tetrad formalism,
via complexifications
$p_A \mapsto p_{A'}$,
$x^\m \mapsto z^{\m\sprime}$.
Consequently, it can be diagrammatically proven that
all Feynman diagrams
for the Kerr spinning-particle
are mere complexifications of those of the Schwarzschild scalar particle.
This implies 
that the amplitudes for $(n{\mem-\,}2)$ gravitons satisfy
\begin{align}
	\label{spin-exp}
	\M_\text{Kerr}(3^+{\cdots\mem}n^+)
	\,=\,
		\M_\text{Sch}(3^+{\cdots\mem}n^+)\,
			\mathe^{(\sum k) \cdot a}
	\,,
\end{align}
where $(\sum k) = k_3 + \cdots + k_n$ is the total momentum.
Here, the asymptotic free trajectory 
is $z^{\m\sprime} = -ia^{\m\sprime} + mu^{\m\sprime}\mem \t$
for constant spin $a^{\m\sprime}$
and velocity $u^{\m\sprime}$ parameters
\footnote{
	We are sorry that our convention for the spin length pseudovector
	flips sign from the standard convention:
	$y^\m = -a^\m$,
	so $S_{12} = \varepsilon_{1203} p^0 a^3 = (+1)\mem (+m)\mem a^3$
	in the rest frame.
}.

It suffices to work in heaven for showing \eqref{spin-exp}
because heaven is by definition made of positive-helicity gravitons
(in the incoming convention).
Also, it should be clear that
any real uplift of the action in \eqref{L}
implies
the all-multiplicity
spin exponentiation
for both positive- and 
negative-helicity gravitons.
One should be able to reproduce the all-multiplicity spin exponentiation with any such real uplift,
as scattering amplitudes are (worldline) field redefinition invariant.

\skip
\paragraph{Uplift to Earth}%
Eventually,
we elaborate on the uplifting procedure,
which provides the extension to generic backgrounds
with \textit{both self-dual and anti-self-dual modes}.
For instance, recall $\a^\m{}_\n$ in \eqref{alpha}
and suppose the Wick rotation in \eqref{wick}
is applied.
On the support of self-duality,
each imaginary unit $i$ 
is absorbed to a Riemann tensor.
Clearly, ambiguities arise from $\O(y^4)$
due to the nonlinear-in-Riemann term:
which Riemann tensor
are we absorbing $i$ into,
$R^\m{}_{\r\s\xi}(x)$ or  $R^\xi{}_{\k\l\n}(x)$?

We dub this ambiguity, in distributing the Hodge stars over nonlinear strings of Riemann tensors,
the \textit{star-shifting ambiguity}.
Its invariant content at 
$\O(R^{n-2})$
in the Lagrangian
can be substantiated as contact deformations of the
$n$-graviton Compton amplitude.

Take $n {\:=\:} 4$, for instance.
By plugging in the straight-line trajectory,
we learn that
it suffices to compare between
the following two terms
in the plane-wave background 
$R_{\m\n\r\s}(x) =
	\e_3\mem \varphi^+_{3\m\n}\mem \varphi^+_{3\r\s}\mem \mathe^{ik_3x}
	+ \e_4\mem \varphi^-_{4\m\n}\mem \varphi^-_{4\r\s}\mem \mathe^{ik_4x}
$,
where
\smash{$\varphi^+_3$} and \smash{$\varphi^-_4$} are gauge-invariant polarizations for the positive- and negative-helicity gravitons:
\begin{subequations}
\label{starshift}
\begin{align}
	\label{starshiftL}
	S^{\m\n}\mem {
		((a{\mem\cdot\mem}\partial)^{j_1}\mem
			{*}R_{\m\n}{}_{a \k})\mem 
		((a{\mem\cdot\mem}\partial)^{j_2}\mem
			R^{\k}{}_{a \r\s})
	}\mem S^{\r\s}
	\,,\\
	\label{starshiftR}
	S^{\m\n}\mem {
		((a{\mem\cdot\mem}\partial)^{j_1}\mem
			R_{\m\n}{}_{a \k})\mem 
		((a{\mem\cdot\mem}\partial)^{j_2}\mem
			{*}R^{\k}{}_{a \r\s})
	}\mem S^{\r\s}
	\,.
\end{align}
\end{subequations}
Upon extracting the part multilinear in $\e_3$ and $\e_4$,
\eqrefs{starshiftL}{starshiftR}
are equal in magnitude but \textit{opposite in sign}:
$
	\pm\mem
	(\varphi^+_3 S)
	(\varphi^-_4 S)
	(
		(k_3a)^{j_1}
		(k_4a)^{j_2}
		{\mem-\mem}
		(k_4a)^{j_1}
		(k_3a)^{j_2}
	)
$ $
	(a\hem\varphi^+_3\varphi^-_4a)
$.
Thus, each time one shifts the Hodge star from one Riemann tensor to another,
one deforms the Compton amplitude
by a contact term
if $j_1 \neq j_2$.

In fact, one can simply add an explicit worldline operator
in the Lagrangian
that involves at least one self-dual and one anti-self-dual curvatures.
Yet, such an explicit deformation may
seem more artificial than star-shifting.

Most generally,
we conclude that the uplift
of the heavenly Kerr EoM
and its Lagrangian
to earth
is ambiguous
due to the possibility of adding such mixed-chirality terms:
\textit{earthly deformations}, so to speak.
Star-shifting is 
a natural instance of earthly deformations.

\skip
\paragraph{Compton Amplitude}%
Note that
star-shifting ambiguities
begin at $\O(y^4)$.
For instance,
we had obtained the spurious-pole-free mixed-helicity gravitational Compton amplitude
from an earthly Kerr Lagrangian \footnote{
	This result was released in
	the talk \cite{IHEStalk25}.
}:
\begin{align}
\begin{split}
\label{MKerr}
&
	\frac{
		\M_\text{Kerr}(3^+4^-)
	}{
		\M_\text{Sch}(3^+4^-)\mem
		\mathe^{-\a/2}\mem \mathe^{\b/2}
	}
\\
&
	=\,
		1 + \gamma + \gamma^2\mem
		\bb{
			\frac{\gamma+\beta}{\alpha+\beta}\mem
			E_2(\a)
			-
			\frac{\gamma-\alpha}{\alpha+\beta}\mem
			E_2(-\b)
		}
	\,,
\end{split}
\end{align}
where $\a,\b,\c$ are linear-in-spin parameters such that
the trivial spin exponentiation is
$
	\M_\text{Sch}(3^+4^-)\mem
	\mathe^{-\a/2}\mem \mathe^{\b/2}
$
and the Arkani-Hamed, Huang, and Huang (AHH) amplitude 
with the spurious pole 
\cite{ahh2017}
is 
$
	\M_\text{Sch}(3^+4^-)\mem
	\mathe^{-\a/2}\mem \mathe^{\b/2}\mem
	\mathe^{\gamma}
$,
and $E_2(\xi) := (\mathe^\xi {\,-\,} \xi {\,-\,} 1)/\xi^2$.
Notably,
\eqref{MKerr}
deviates from AHH
already from $\O(\gamma^4)$,
although the strict necessity of contact deformations 
arises from $\O(\gamma^5)$
\cite{ahh2017}
\footnote{
	A similar phenomenon had occurred in the electromagnetic analog
	\cite{Kim:2024grz}.
}.

\skip
\paragraph{Interpretation}%
The NJA had been in a mysterious status
for decades.
Despite initial criticisms
\cite{ernst1968new2,Drake:1998gf},
constant attempt was made for explanations
\cite{talbot1969newman,Drake:1998gf,Gurses:1975vu,flaherty1976hermitian,grg207flaherty,Rajan:2016zmq,giampieri1990introducing,Newman:1965tw-janis,penrose1967twistoralgebra}.
Especially,
Newman himself
envisioned a fundamental unification of spin and spacetime
into a complex geometry,
in which the spinning particle
truly becomes a scalar particle
\cite{%
	newman1988remarkable,Newman:1973afx,Newman:2002mk,Newman:1973yu,newman1974curiosity,newman1974collection,Newman:1973afx,Newman:2004ba,
	Newman:1976gc,ko1981theory,grg207flaherty%
}.
This idea was not only pursued at the level of sourced fields
but also at the level of probe particles:
``complex center of mass.''
An attempt for EoM
appears in \rcite{ko1981theory}.

This paper
vividly realizes Newman's dream
in terms of 
explicit and rigorous frameworks
of geodesic deviation
\cite{gde}
and
adapted complex structure
\cite{guillemin1991grauert,guillemin1992grauert,lempert1991global,szHoke1991complex,halverscheid2002complexifications,aguilar2001symplectic,burns2000symplectic,hall2011adapted}.
The holomorphic worldline $z^{\m\sprime}$
is Newman's complex center of mass,
realized in fully nonlinear gravity.
However,
our analysis on earthly deformations
clarifies that
Newman's proposal
of the Kerr spinning particle
being equivalent to a scalar particle
is valid only in the self-dual sector.
To reiterate,
the probe NJA
pinpoints a unique dynamics
\textit{within} heaven.
This point is made further evident in the appendix,
where 
an explicit anti-self-dual group of terms
ruins
the equivalence between \eqrefs{Lz}{L} in earth.

A similar realization has been critical in
grasping the full content of the NJA \cite{nja}.
The self-dual limit of
the Kerr-Taub-NUT solution
is diffeomorphic to the self-dual Taub-NUT solution \cite{Crawley:2021auj}.
In Kerr-Schild coordinates \cite{note-sdtn,nja},
this diffeomorphism is just a translation by $+i\vea$.
Crucially, however,
the non-self-dual Kerr solution
describes \textit{two} centers at $\pm i\vea$
in the complexified description:
a pair of self-dual \textit{and anti-self-dual} Taub-NUT solutions
\cite{nja}.
Thus, the equivalence between
Kerr and a pointlike object
is lost
beyond the self-dual sector.

Based on this repertoire
that applies for both our probe NJA and the NJA,
it seems that a natural interpretation for the
holomorphic worldline $z^{\m\sprime}$ in \eqref{z}
is the worldline of the self-dual Taub-NUT instanton
constituting the Kerr black hole.
Surely,
the anti-holomorphic worldline could have also been
obtained by
geodesically deviating in the direction of $-iy^\m$,
in which case the worldline of the anti-self-dual Taub-NUT instanton
is reached.

This interpretation is nicely consistent with spin exponentiation,
disclosing an underlying physical picture:
spin exponentiation is the helicity selection rule \cite{Adamo:2023fbj,Adamo:2024xpc,Adamo:2025fqt} of Taub-NUT instantons.
It also provides a physical interpretation for the worldsheet structure speculated in \rcite{gmoov}:
the Misner string
(gravitational Dirac string)
as a flux tube of NUT charge
joining the self-dual and anti-self-dual instantons.

Due to the nonlinearity of gravity
inherent in \eqref{z},
there is a difficulty in
simultaneously manifesting the nonlinear Newman-Janis shifts
(all-multiplicity spin exponentiations, spin resummations)
for both self-dual (positive-helicity) and anti-self-dual (negative-helicity) sectors.
Still, we shall make
maximum use of the superintegrability of the self-dual sector
shown in \eqref{K}
by perturbing around the holomorphic variable $z^{\m\sprime}$:
a \textit{googly}
\cite{Penrose:2015lla:Palatial,penr04-googly,Witten:2003nn,Adamo:2017qyl}
approach
to Kerr spinning-particle effective theory.

\skip
\paragraph{Conclusion}%
We proposed
the probe counterpart of NJA
to uniquely pinpoint
a remarkable instance of a massive spinning particle
in self-dual spacetimes,
which exhibits exact symmetries
in the presence of Killing vectors and Killing-Yano tensors.
It is shown that this distinguished point
in the effective theory landscape of all massive spinning particles
enjoys
the privileges of
superintegrability in self-dual Taub-NUT background
and
spin exponentiation at every graviton multiplicity.
It is also determined that the uplift to real spacetimes
is ambiguous 
precisely by mixed-chirality deformations,
corresponding to contact deformations for Compton amplitudes.

This paper provides a partial answer to the persistent question,
``What principle in the effective theory
characterizes Kerr among all massive spinning objects?''
Previously, it has been speculated as
a \textit{simplicity} of linearized interactions
\cite{ahh2017,Guevara:2018wpp,Guevara:2019fsj,chkl2019,aho2020}
or its continuation \cite{Johansson:2019dnu,Aoude:2020onz,Lazopoulos:2021mna},
a \textit{pattern} suggested at low orders \cite{Bern:2022kto,Aoude:2022trd},
or the \textit{gauge redundancy} in the higher-spin formalism \cite{Ochirov:2022nqz,Cangemi:2022bew}.
However, we identified 
an exact hidden \textit{symmetry}
embedded in nonlinear interactions.
For the full non-self-dual dynamics,
a googly search for earthly deformations
that breaks 
superintegrability in self-dual Taub-NUT
\textit{partially}
down to integrability in Kerr
might be a reasonable direction.

\medskip
\paragraph{Acknowledgements}%
We would like to thank
Thibault Damour, Julio Parra-Martinez, and Justin Vines
for many stimulating discussions.
J.-H.K. is supported by the Department of Energy (Grant No.~DE-SC0011632) and by the Walter Burke Institute for Theoretical Physics.

\newpage

\pagebreak
\appendix
\onecolumngrid
	\begin{center}
	\bfseries\normalsize Appendix
\end{center}
\renewcommand{\theequation}{A.\arabic{equation}}
\setcounter{equation}{0}

\vspace{6pt}
\noindent\textbf{Flat.}
There exists a variety of options for the Lagrangian formulation of a massive spinning particle:
bivector implementation \cite{Hanson:1974qy,bailey1975lagrangian},
pseudovector implementations \cite{ambikerr0,Scheopner:2023rzp},
minimal implementation \cite{souriau1974modele},
etc.
However,
it should be emphasized that
one's choice of a formulation should not affect
the EoM for physical degrees of freedom.
Hence,
for the purpose of this appendix,
it simply suffices to state just one well-defined instance \cite{ambikerr0}:
\begin{align}
	\label{theta0}
	S_0[x,y,\lambda,\rambda;\k^0,\k^1]
	\,=
	\int\mem
		p_\m\mem dx^\m
		+ p_\m\mem {*}\Theta^\m{}_\n\mem y^\n
		- \mathscr{C}
	\quad\text{where}\quad
	\mathscr{C}
	\,=\,
		\frac{\smash{\k^0}}{2}\mem \BB{
			p^2 + m^2
		}\, d\t
		+ \smash{\k^1}\mem \BB{
			-p\hem \mdot y
		}\, d\t
	\,.
\end{align}
Here, $\lambda_\a{}^I \in \mathrm{GL}(2,\mathbb{C})$
and $\rambda_{I\da} = [\lambda_\a{}^I]^*$
are complex variables such that
\begin{align}
	p_{\a\da}
	\,=\,
		-\lambda_\a{}^I \rambda_{I\da}
	\,,\quad
	p_\m\mem {*}\Theta^\m{}_\n\mem y^\n
	\,=\,
		i\hem y^{\da\a}\mem 
		\BB{
			\lambda_\a{}^I\mem D\rambda_{I\da}
			-
			D\lambda_\a{}^I\mem \rambda_{I\da}
		}
	\,.
\end{align}
Here, $I = 0,1$ is a global $\mathrm{SU}(2)$ index;
$\k^0$ and $\k^1$ are Lagrange multipliers;
$D$ denotes the covariant exterior derivative
with respect to the Levi-Civita connection of the spacetime metric $g$,
which we are taking as flat for the moment.

\eqref{theta0}
describes a massive spinning particle
with three translation and three spinning degrees of freedom,
which is shown \cite{ambikerr0} to be equivalent to the widely-adopted model \cite{Hanson:1974qy}
up to double cover.
The covariant spin supplementary condition 
$S^{\m\n} p_\n = 0$ is automated via
$S^{\m\n} = \ve^{\m\n\r\s} y_\r p_\s$,
while $\k^1$ imposes $p\hem \mdot y = 0$
as a constraint
to achieve \eqref{ydef}.


\vspace{6pt}
\noindent\textbf{Earth.}
Now take the spacetime 
be a real-analytic four-manifold
equipped with a pseudo-Riemannian metric $g_{\m\n}$
whose Riemann tensor $R^\m{}_{\n\r\s}$ is nonvanishing.
Define the so-called $Q$-tensors \cite{gde} as
\begin{align}
	\label{def-Qtens}
	(Q_\ell)^\m{}_\s
	\,:=\,
		y^{\k_1}{\cdots}y^{\k_\ell}\mem
		R^\m{}_{\k_1\k_2\s;\k_3;\cdots;\k_\ell}\hnem(x)
	\,,\quad
	({*}^\ell Q_\ell)^\m{}_\s
	\,:=\,
		y^{\k_1}{\cdots}y^{\k_\ell}\mem
		\mem
		{*}^\ell R^\m{}_{\k_1\k_2\s;\k_3;\cdots;\k_\ell}\hnem(x)
	\,,
\end{align}
where ${*}^\ell$ means to act on the Hodge dual
$\ell$ times.
Then 
a natural earthly uplift of \eqref{L} is
$\int \theta - \mathscr{C} $,
where
\begin{subequations}
\begin{align}
    \label{theta-Kerr-earth}
    \theta
    \,=\,
    {}&{}
   	p_\m\mem dx^\m
   	\,+\,
   		p_\m\mem {*}\Theta^\m{}_\n\mem y^\n
    \,+\,
	    \sum_{\ell=2}^\infty\mem
	        \frac{\hem{1}\hhem}{\ell!}\,
	        p_\m\hem
	        \BB{
	        	(\i_N D)^{\ell-2}\mem \i_N\, {*}^\ell R^\m{}_\n
	        }
	        \hem y^\n
    \,,\\
\begin{split}
    \label{kerrEFT.quadratic}
    \,=\,
    {}&{}
     	p_\m\mem dx^\m
     	\,+\,
     		p_\m\mem {*}\Theta^\m{}_\n\mem y^\n
    \\
    {}&{}
    + 
    \sum_{\ell=2}^\infty\,
        \frac{1}{\ell!}\,\mem
        p_\m\hhnem
        \bigg[\mem{
            ({*}^\ell Q_{\ell})^\m{}_\s \mem{dx^\s}
            + (\ell{\mem-\mem}2)\mem ({*}^\ell Q_{\ell-1})^\m{}_\s \mem{Dy^\s}
        }\mem\hem\bigg]
    \\
    {}&{}
    + 
    \sum_{\ell=2}^\infty\,
        \frac{1}{\ell!}\,
    \mem
    \sum_{j=2}^{\ell-2}\,
    \binom{\ell{\mem-\mem}2}{j}\,\mem
    p_\m\hhnem
    \bigg[\mem{
        ({*}^\ell Q_{\ell-j} Q^{\vphantom{+}}_{j})^\m{}_\s \mem{dx^\s}
        + (j{\mem-\mem}2)\mem ({*}^\ell Q_{\ell-j} Q^{\vphantom{+}}_{j-1})^\m{}_\s \mem{Dy^\s}
    }\mem\hem\bigg]
    + \mathcal{O}(R^3)
    \,.
\end{split}
\end{align}
\end{subequations}
Writing down the complete formula to all orders in spin and curvature
is left as a trivial exercise \cite{gde}.
\eqref{theta-Kerr-earth}
summons the $\i_N D$ formalism of \rcite{gde},
which is not necessary for comprehending \eqref{kerrEFT.quadratic}.
Namely, $\i_N$ denotes the interior product with respect to $N$,
while
$N$ is the horizontal lift \cite{ehresmann1948connexions}
implementing the generator of geodesic deviation,
defined as the unique vector field on the phase space
such that
\begin{align}
	\i_N dx^\m \,=\, y^\m
	\,,\quad
	\i_N Dy^\m \,=\, 0
	\,,\quad
	\i_N D\lambda_\a{}^I \,=\, 0
	\,,\quad
	\i_N D\rambda_{I\da} \,=\, 0
	\,.
\end{align}

\vspace{6pt}
\noindent\textbf{Heaven.}
In \eqref{theta-Kerr-earth},
$R^\m{}_\n = \frac{1}{2}\mem R^\m{}_{\n\r\s}(x)\mem dx^\r \swedge dx^\s$ denotes the curvature two-form.
We take ${*}R^\m{}_\n = i\mem R^\m{}_\n$ to enter heaven 
and find
\begin{align}
	\label{heavenly portal}
    \theta
   	\,&=\,
 	   	\mathe^{i\pounds_N}\mem \BB{\mem
 	   		p_\m\mem dx^\m
	 	   	\,+\,
	 	   		p_\m\mem {*}\Theta^\m{}_\n\mem y^\n
	 	   	\,-\,
	 	   		i\mem p_\m Dy^\m
	   	\hem}
	\,=\,
  		p_\m\mem dz^{\m\sprime}
	   	\,+\,
	   		p_{\m'}\mem {*}\Theta^{\m\sprime}{}_{\n'}\mem y^{\n\sprime}
	   	\,-\,
	   		i\mem p_{\m'} Dy^{\m\sprime}
	\,,
\end{align}
where we have used the fact that
$p_\m\mem {*}\Theta^\m{}_\n\mem y^\n - i\mem p_\m Dy^\m$
is ``Newman-Janis shift invariant'' in heaven
(recall \eqref{D=0}).
\eqref{heavenly portal} establishes the resummation of the action
on the holomorphic worldline.
$\pounds_N$ denotes the Lie derivative.
The constraints in \eqref{theta0} are scalars
and thus are trivially brought to the holomorphic worldline.

Meanwhile, the free theory of the spinning particle
is described by the symplectic potential
$\theta^\bullet = p_\m\mem dx^\m + p_\m\mem {*}\Theta^\m{}_\n\mem y^\n$,
as is shown in \eqref{theta0}.
Their precise relation implies
\begin{align}
	\mathe^{-i\pounds_N}
	\theta
	\,=\,
		\theta^\bullet
		\,-\,
			i\mem p_\m\mem Dy^\m
	\qiq
	\mathe^{-i\pounds_N} \omega
	\,=\,
		\omega^\bullet
		\,-\,
			i\mem Dp_\m \swedge Dy^\m
	\,,
\end{align}
where $\omega^\bullet$ is a two-form such that $\omega = d\theta$.
As a result, the Hamiltonian EoM arise from the vector field
\begin{align}
	\omega^{-1}(\blank,D\phi_i)
	\,=\,
		\omega^\bullet{}^{-1}(\blank,D\phi_i)
		\mem+\mem
		i\mem
			\omega^\bullet{}^{-1}(\blank,Dp_\m)\,
			\omega^\bullet{}^{-1}(Dy^\m,D\phi_i)
		\mem-\mem
		i\mem
			\omega^\bullet{}^{-1}(\blank,Dy^\m)\,
			\omega^\bullet{}^{-1}(Dp_\m,D\phi_i)
		\mem+\mem
		\cdots
	\,,
\end{align}
where $\phi_i$ collectively denotes the first-class constraints $\phi_0 = (p^2 \mplus m^2)/2$ and $\phi_1 = -p\hem\mdot y$.
It is not difficult to see that
both
$\omega^\bullet{}^{-1}(Dy^\m,D\phi_i)$
and
$\omega^\bullet{}^{-1}(Dp_\m,D\phi_i)$
vanish
due to an isomorphism with free theory Poisson brackets,
so
\begin{align}
	\label{ioms}
	\omega^{-1}(dx^\m,D\phi_i)\, \k^i
	\,=\,
		p^\m\mem \k^0
	\,,\quad
	\omega^{-1}(Dp_\m,D\phi_i)\, \k^i
	\,=\,
		0
	\,,\quad
	\omega^{-1}(Dy^\m,D\phi_i)\, \k^i
	\,=\,
		0
	\,,
\end{align}
where sum over $i=0,1$ is implied.
By gauge-fixing the einbein as $\k^0 = 1$
and acting on $\mathe^{i\pounds_N}$ to \eqref{ioms},
one readily reproduces \eqref{heaven.z}.
This establishes the derivation of \eqref{heaven.z}
from the Lagrangian in \eqref{Lz}.

\vspace{6pt}
\noindent\textbf{Spin Precession.}
By using the identities 
$Dp^{\m\sprime} = W^{\m\sprime}{}_\m\mem (\mathe^{\pounds_N^D} Dp^\m)$
and
$Dy^{\m\sprime} = W^{\m\sprime}{}_\m\mem (\mathe^{\pounds_N^D} Dy^\m)$
\cite{gde},
\eqref{heaven.z} derives that
the antisymmetric tensor $\Omega^\m{}_\n$ in \eqref{heaven}
is given by the following
(the acute-$Q$-tensors in \rcite{gde}):
\begin{align}
\label{xplc}
	&
	\frac{1}{1!}\,
	\bb{
		y^\r
		\mem{*}R^\m{}_{\n\r\s}(x)\mem
			\frac{dx^\s}{d\t}
	}
	-
	\frac{1}{2!}\,
	\bb{
		y^\r y^\k 
		R^\m{}_{\n\r\s;\k}(x)\mem
			\frac{dx^\s}{d\t}
		+
		y^\r
		R^\m{}_{\n\r\s}(x)\mem
			\frac{Dy^\s}{d\t}
	}
	\\
	&-
	\frac{1}{3!}\,
	\bb{
		y^\r y^{\k_1} y^{\k_2} 
		\mem{*}R^\m{}_{\n\r\s;\k_1;\k_2}(x)\mem
			\frac{dx^\s}{d\t}
		+
		y^\r y^{\k_1} y^{\k_2}
		\mem{*}R^\m{}_{\n\r\l}(x)\mem
			R^\l{}_{\k_1\k_2\s}(x)\mem
			\frac{dx^\s}{d\t}
		+
		2\mem 
		y^\k y^\r
		\mem{*}R^\m{}_{\n\r\s;\k}(x)\mem
			\frac{Dy^\s}{d\t}
	}
	+ \cdots
	\,.
	\nonumber
\end{align}
Here, we have traded off $i$ with $*$
to be instructive.
It is a nice exercise to check that
the $\O(R^1)$ part of \eqref{xplc}
derives 
the all-orders-in-spin completion of
the Mathisson-Papapetrou-Dixon equations \cite{Mathisson:1937zz,Papapetrou:1951pa,Dixon:1970zza}
in the covariant spin supplementary condition,
for the Levi-Steinhoff \cite{Levi:2015msa} multipole coefficients
$C_\ell = 1$.
The universal part
describes the gravitational gyromagnetic ratio 
$g=2$
\cite{Mathisson:1937zz,Papapetrou:1951pa,Dixon:1970zza,chkl2019};
the quadrupolar extension describes 
the gravimagnetic ratio $\kappa=1$ \cite{Yee:1993ya,khriplovich1989particle,Khriplovich:1997ni};
the all-multipole extension
is described in Appendix A of \rcite{ambikerr1}.

\vspace{6pt}
\noindent\textbf{Googly View on Earth.}
Finally, note that
\begin{align}
    \theta^\pm_1
    \,=\,
    \frac{1}{2}\,
    p_\m\mem
    \Big({
        Dy^m
        \mp i\mem {*\Theta}^\m{}_\n\mem y^\n
    }\Big)
    \qiq
    (\pounds_N)^{\ell-1}
    \theta^\pm_1
    =
    p_\m\hem
    (\hhnem(\i_N D)^{\ell-2}\mem \i_N{R^\pm}{}^\m{}_\n\hhem)
    \hem y^\n
    \,,
\end{align}
where ${R^\pm}{}^\m{}_\n = \frac{1}{2}\mem (1 \mp i\mem{*})R^\m{}_\n$
are the self-dual and anti-self-dual curvature two-forms.
In turn,
it follows that
\begin{align}
    \theta
    \,-\,
      	\mathe^{i\pounds_N}\BB{
      		p_\m\mem dx^\m
      	}
     \,=\,
    \frac{\hem{
        \mathe^{-i\pounds_N}
        {\mem-\mem}\mem
        \mathe^{i\pounds_N}
    }\hhem}{\pounds_N}\mem
    \theta^-_1
    \,,
\end{align}
which computes ``$\text{Kerr} - (\text{Scwharzschild $+iy$})$.''
As a result, we find
\begin{align}
    \label{hol-earth-kerr}
    \theta
    \,&=\,
    \mathe^{i\pounds_N}
    \bigg(\hem\hhhem{\,
        p_\m\hem dx^\m
        +
        \frac{\hem{
            \mathe^{-2i\pounds_N}
            {\mem-\mem}
            1
        }\hhem}{\pounds_N}\mem
        \theta^-_1
    \,}\hhhnem\bigg)
    \,,\\
    \,&=\,
    \mathe^{i\pounds_N}
    \bigg(\hem\hhhem{\,
        p_\m\hem dx^\m
        + p_\m\mem {*\Theta}^\m{}_\n\mem y^\n
        - i\mem p_\m\hem Dy^\m
        +
        \sum_{\ell=2}^{\infty}
        \frac{(-2i)^{\ell}}{\ell!}\mem
        p_\m\hem
        (\hhnem(\i_N D)^{\ell-2}\mem \i_N{R^-}{}^\m{}_\n\hhem)
        \hem y^\n
    \,}\hhhnem\bigg)
    \,.
    \nonumber
\end{align}
Distributing \smash{$\mathe^{i\pounds_N}$},
all the tensors here get analytically continued and evaluated at \smash{$z^{\m\sprime}$}.
This derives that
\begin{align}
    \label{googly}
    \theta
    \,=\,
    {}&{}
    \bb{
     	p_{\m'}\mem dz^{\m\sprime}
     	\,+\,
     		p_{\m'}\mem {*}\Theta^{\m\sprime}{}_{\n'}\mem y^{\n\sprime}
     	\,-\,
     	i\mem p_{\m'}\mem Dy^{\m\sprime}
    }
    \,+\,
    \sum_{\ell=2}^\infty\,
        \frac{(-2i)^\ell}{\ell!}\,\mem
        p_{\m'}\hhnem
        \bigg[\mem{
            (Q_{\ell}^-)^{\m\sprime}{}_{\s'} \mem{dz^{\s\sprime}}
            + (\ell{\mem-\mem}2)\mem (Q_{\ell-1}^-)^{\m\sprime}{}_{\s'} \mem{Dy^{\s\sprime}}
        }\mem\hem\bigg]
    \\
    {}&{}
    +\,
    \sum_{\ell=2}^\infty\,
        \frac{(-2i)^\ell}{\ell!}\,
    \mem
    \sum_{j=2}^{\ell-2}\,
    \binom{\ell{\mem-\mem}2}{j}\,\mem
    p_{\m'}\hhnem
    \bigg[\mem{
        (Q_{\ell-j}^- Q^{\vphantom{+}}_{j})^{\m\sprime}{}_{\s'} \mem{dz^{\s\sprime}}
        + (j{\mem-\mem}2)\mem (Q_{\ell-j}^- Q^{\vphantom{+}}_{j-1})^{\m\sprime}{}_{\s'} \mem{Dy^{\s\sprime}}
    }\mem\hem\bigg]
    \nonumber
    \\
    {}&{}
    +\,
    \sum_{\ell=2}^\infty\,
        \frac{(-2i)^\ell}{\ell!}\,
    \mem
    \sum_{j_1=2}^{\ell-4}\mem
    \sum_{j_2=2}^{\ell-4}\,
    \binom{\ell{\mem-\mem}2}{j_1+j_2}\mem
    \binom{j_1{\mem+\mem}j_2{\mem-\mem}2}{j_2}\,\mem
    p_{\m'}\hhnem
    \bigg[\mem{
        (Q_{\ell-j_1-j_2}^- Q^{\vphantom{+}}_{j_1} Q^{\vphantom{+}}_{j_2})^{\m\sprime}{}_{\s'} \mem{dz^{\s\sprime}}
        + (j_2{\mem-\mem}2)\mem (Q_{\ell-j_1-j_2}^- Q^{\vphantom{+}}_{j_1} Q^{\vphantom{+}}_{j_2-1})^{\m\sprime}{}_{\s'} \mem{Dy^{\s\sprime}}
    }\mem\hem\bigg]
    \nonumber
    \\
    {}&{}
    + \mathcal{O}(R^4)
    \,,
    \nonumber
\end{align}
yielding the googly formulation of Kerr Lagrangian to all orders.
Here, $Q^-_\ell := \frac{1}{2}\mem ( Q_\ell + i\, {*}Q_\ell )$
with primed indices are composed of
(covariant derivatives of)
$R^{\m\sprime}{}_{\n\sprime\mem\r\sprime\mem\s\sprime\,}(z)$
and $y^{\k\sprime}$.
Crucially, \eqref{googly} equivalently rewrites the completely \textit{real} Lagrangian
due to \eqref{kerrEFT.quadratic}
in terms of holomorphic, complexified variables.
such as
$z^{\m\sprime}$, $p_{\m'}$, and $y^{\m\sprime}$.
The first line in \eqref{googly} is the linear coupling.
From the second line, however, each term is $\O((R^+)^{n_+}(R^-)^{n_-})$
with $n_- \geq 1$.

Consequently,
\eqref{googly} 
shows that an earthly Kerr Lagrangian
equals the heavenly Kerr Lagrangian in \eqref{Lz} plus anti-self-dual curvature corrections,
which substantiates the point in the main article.

\pagebreak
\twocolumngrid

\bibliography{references.bib}

\end{document}